\documentclass[doublecol]{epl2}

\usepackage{amsmath,amssymb,bm,bbold}

\usepackage[integrals]{wasysym}

\title{Instability of sliding Luttinger liquid}

\author{V. Fleurov\inst{1}, V. Kagalovsky\inst{2,5},  I. V. Lerner\inst{3} \and  I. V. Yurkevich\inst{4,5}}
\shortauthor{V. Fleurov \etal}

\institute{
\inst{1} Raymond and Beverly Sackler Faculty of Exact Sciences,
School of Physics and Astronomy,
Tel-Aviv University -  Tel-Aviv 69978, Israel\\
\inst{2} Shamoon College of Engineering - Beer-Sheva 84105, Israel\\
\inst{3} University of Birmingham, School of Physics \& Astronomy -   Birmingham B15 2TT, UK\\
\inst{4} Aston University, School of Engineering \& Applied Science -  Birmingham B4 7ET, UK\\
\inst{5} Institute of Fundamental and Frontier Sciences, University of Electronic Science and Technology of China -  Chengdu 610054, People's Republic of China
}

\abstract{We revise a phase diagram for the sliding Luttinger liquid (SLL) of coupled one-dimensional quantum wires packed in two- or three-dimensional arrays. We analyse whether   physically justifiable (reasonable) inter-wire interactions, i.e.\ either the screened Coulomb or  ``Coulomb-blockade" type interactions, stabilise the SLL phase. Calculating the scaling dimensions of the most relevant perturbations (the inter-wire single-particle hybridisation, charge-density wave, and superconducting inter-wire couplings), we find that their combination   always destroys the SLL phase for  the repulsive intra-wire  interaction. However,  suppressing the inter-wire tunnelling (when the charge-density wave is the only remaining perturbation), one can observe a stability region emerging due to the inter-wire interaction.}

\begin{document}

\maketitle

\def\e{\mathrm{e}}

\section{Introduction}

The Luttinger liquid (LL) describes one-dimensional interacting systems with a linear (or linearised) spectrum \cite{Tomo,Lutt:63,Haldane:81}. The interaction strongly enhances the impact of impurities leading to a zero-temperature metal-insulator transition in the presence of either disorder \cite{GS1988}  or even a single impurity \cite{KaneFis:92a,FurusakiNagaosa:93b}, both being described by the renormalisation group (RG) approach.
Progress in the fabrication of low-dimensional nanostructures based on carbon nanotubes \cite{Bockrath:01,Ishii:03,CBKCCL04,Mason:04}, semiconductor and metallic nanowires \cite{MBDGBPH03,SHZZ04}, self-assembled DNA scaffolds \cite{S03,S09}, etc., revived the interest to theoretical studies of the LL superstructures. Theoretically, the most challenging bunch of problems in this field is the crossover from the 1D LL behaviour of an individual metallic wire   to a 2D or 3D Fermi-liquid (FL) behaviour of ensembles of coupled 1D wires. Such finite systems may support various topological states with gapless modes propagating through the edge wires \cite{NCMT14}.

It is known that the inter-wire electron tunnelling is a relevant perturbation which results in the transition from the LL to FL phase \cite{W90,Giamarchi}. Nevertheless, the LL fixed point remains (under proper conditions) stable in a phase of the sliding Luttinger liquid (SLL) \cite{GG98,OHern1999,EFKL00,MKL2001,Kane2002,Vishwanath2001}. In this phase,  the canonic phase -- density variables $(\varphi_j, \theta_j)$ describing bosonised degrees of freedom in each wire $j$ are invariant under the constant shifts. The phase remains stable as long as  three distinct inter-wire processes are all RG irrelevant. Namely, these processes are
the  single-particle inter-wire tunnelling (SP), the particle-hole hopping that may result in a transverse charge density wave (CDW), and the two-particle hopping resulting in a superconducting  (SC) state.   The SP perturbation becomes irrelevant when a spin gap appears due to an intra-wire large-angle spin-flip scattering  relevant for the positive Luttinger parameter, $K_0 > 0$, in a single wire, or for $K(q_\bot) >0$ in an array  (the Luther-Emery regime \cite{LE74,Giamarchi,SL05}). Then  the stability of the SLL fixed point would be ensured if the remaining SC and CDW perturbations also become  irrelevant \cite{MKL2001,Kane2002}.

In the region where  $K(q_\bot) < 0$ the large-angle scattering is irrelevant, the gap does not appear and we deal with a spinful LL unless an in-plane magnetic field is applied \cite{Vishwanath2001}. Then the system becomes effectively spinless, but the SP processes are not necessarily irrelevant. Then all three types of interaction, CDW, SC, and SP, must be simultaneously irrelevant in order to ensure the stability of SLL. As shown in \cite{Vishwanath2001} this can take place in a narrow stripe on the margins of the allowed parameter range. A special shape of the transverse wave vector dependence of the Luttinger parameter is used, whose microscopic derivation is not obvious. Moreover, it is quite clear that the result is extremely sensitive to the choice of the shape of this dependence and the values of the parameters.

In this letter we  study whether long-range inter-wire interactions ({modelled either as a screened Coulomb interaction, or a distance-independent `Coulomb-blockade' type interaction in a finite array})  can stabilise the SLL phase in a system of coupled parallel quantum wires packed in two- and three-dimensional arrays. We show that this is not the case for fermionic wires (with the Luttinger parameter $K\le1$): all the three important perturbations cannot be made simultaneously irrelevant for these types of long-range interactions. On the other hand, we show if the direct inter-wire tunnelling is suppressed then there is a region of parameters where the only remaining (for achievable temperatures)  CDW perturbation becomes irrelevant and   that the SLL phase becomes stable.

\section{The model}
After the standard bosonisation \cite{Giamarchi}, the density fluctuations and current in the $i^{{\mathrm{th}}} $ wire are parameterised in terms of two  bosonic fields $\theta_i$ and $\varphi _i$ as
$
\delta\rho_i =\frac{1}{\pi}\partial_x\theta_i, $ and $ j_i=\frac{1}{\pi}\partial_x\varphi_i.
$
Introducing the vector notations for the two fields describing  a set of $N$ wires,
\begin{align}
{\bm\theta}& = \{\theta_1,...,\theta_N\},&
{\bm\varphi} &= \{\varphi_1,...,\varphi_N\},
\label{phitheta}
\end{align}
one writes the Lagrangian density of the set as
\begin{equation}\label{LN}
{\cal L}=\frac{1}{8\pi} \left[2\partial_t {\bm\varphi}^{\rm T} \partial_x{\bm\theta}
-\partial_x{\bm\theta}^{\rm T} {\mathsf{V}}_{\theta} \partial_x{\bm\theta}-\partial_x{\bm\varphi}^{\rm T} {\mathsf{V}}_{\varphi} \partial_x{\bm\varphi}\right] .
\end{equation}
Here the matrices ${\mathsf{V}}_\theta$ and ${\mathsf{V}}_\varphi $ are diagonal in the absence of inter-wire interactions, with the elements expressed in terms of  the velocity $v_i$ and the Luttinger parameter $K_i$ in each wire as $V_\theta^{ij}=\delta_{ij}(v_i/K_i)  $ and $V_\varphi ^{ij}=\delta_{ij} v_i K_i   $. Adding the inter-wire interactions makes these matrices non-diagonal,
\begin{align}\label{V}
V^{ij}_{\theta}&=\delta_{ij}(v_i/K_i)+U_{\theta}^{ij}\,,& V^{ij}_{\varphi}&=\delta_{ij} v_i K_i +U_{\varphi}^{ij}\,,
\end{align}
where the off-diagonal
 matrix elements $U_{\theta}^{ij}$ and $U_{\varphi }^{ij}$ describe the density-density and current-current interaction strengths between the  $i^{{\mathrm{th}}} $ and $j^{{\mathrm{th}}} $ wires.

In the presence of the inter-wire interaction, the local field correlators with Lagrangian density \eqref{LN} can be written in matrix form as
\begin{align}\label{corr}
\begin{aligned}
\langle{\bm\theta}(t)\otimes{\bm\theta}^{\rm T}(t')\rangle&=-2{\mathsf{K}}\,\ln (t-t')\,,\\
\langle{\bm\varphi}(t)\otimes{\bm\varphi}^{\rm T}(t')\rangle&=-2{\mathsf{K}}^{-1}\,\ln (t-t')\,,
\end{aligned}
\end{align}
where the \emph{Luttinger matrix} ${\mathsf{K}}$ is defined \cite{YGYL:2013,IVY:2013,KagLY} by
  the matrix equation
\begin{equation}
{\mathsf{K}}\,{\mathsf{V}}_{\theta}\,{\mathsf{K}}= {\mathsf{V}}_{\varphi}\,,
\label{Kmat}
\end{equation}
which always
has a unique solution for real symmetric and positive definite matrices ${\mathsf{
V}}_\theta$ and ${\mathsf{V}}_\varphi $. The name is justified by the fact that, in the absence of inter-wire interactions, ${\mathsf{K}}=\operatorname{diag}\{{K_i}\} $, and Eqs.~\eqref{corr} are reduced to the standard single-wire expressions.

\section{Perturbations}
It is convenient to write operators corresponding to the three inter-wire processes defined in the introduction (SP, CDW and SC) in terms of the creation and annihilation operators for the left-moving ({$\hat{L}_i^\dagger  $ and $\hat{L}_i$}) and right-moving ({$\hat{R}_i^\dagger  $ and $\hat{R}_i$}) particles in the $i^{{\mathrm{th}}}  $ wire, with $\hat{L}_i\sim \e^{i\theta^L_i} $ and  $\hat{R}_i\sim \e^{i\theta^R_i} $, where the bosonic variables for the left- and right-movers are $\theta^{L,R}_i\equiv \varphi _i\pm \theta_i $.
Then
the matrix elements of the three potentially relevant
inter-wire couplings are given by
\begin{align}\label{inter}
\begin{aligned}
{\hat L}_{ij}^{\rm cdw}&\sim    {\hat R}^{\dagger}_i{\hat L}_i\,{\hat L}^{\dagger}_j{\hat R}_j\sim\cos\left[\theta_i-\theta_j\right]\,,\\
{\hat L}_{ij}^{\rm sc}&\sim   {\hat R}^{\dagger}_i {\hat L}^{\dagger}_i\,{\hat L}_j{\hat R}_j\sim\cos\left[\varphi_i-\varphi_j\right]\,,\\
{\hat L}_{ij}^{\rm sp}&\sim   {\hat R}^{\dagger}_i{\hat R}_j\sim \e^{i(\theta_i-\theta_j)/2}\,\e^{i(\varphi_i-\varphi_j)/2}\,.
\end{aligned}
\end{align}
The corresponding scaling dimensions are straightforward to derive  using field correlators \eqref{corr}; the results are expressed in terms of matrix elements of the Luttinger matrix ${\mathsf{K}}$ and its inverse ${\mathsf{K}}^{-1} $:
 \begin{align}
 \begin{aligned}
 \Delta_{ij}^{\rm cdw}&\equiv{\rm dim}\left[{\hat L}_{ij}^{\rm cdw}\right]=K_{ii}+K_{jj}-2K_{ij}\,,\\
\Delta_{ij}^{\rm sc}&\equiv{\rm dim}\left[{\hat L}_{ij}^{\rm sc}\right]=({\mathsf{K}}^{-1})_{ii}+({\mathsf{K}}^{-1})_{jj}-2({\mathsf{K}}^{-1})_{ij}\,,\\
\Delta_{ij}^{\rm sp}&\equiv{\rm dim}\left[{\hat L}_{ij}^{\rm sp}\right]=\frac{1}{4}\,\left[\Delta_{ij}^{\rm cdw}+\Delta_{ij}^{\rm sc}\right]\,.
\end{aligned}\label{scal1}
 \end{align}

The stability condition (i.e.\ the irrelevance of all the three perturbations) for a one-dimensional system is that all the three scaling dimensions  are greater than the physical dimension, $1+1=2$:
\begin{align}\label{dim}
\Delta_{ij}^{\rm cdw}&\geq   2\,,& \Delta_{ij}^{\rm sc}&\geq 2\,,&
\Delta_{ij}^{\rm cdw} + \Delta_{ij}^{\rm sc}&\geq 8\,.
\end{align}
The last inequality is potentially most stringent so that the single-particle hybridisation might be  dangerous for the stability of the SLL phase even when both CDW and SC processes are irrelevant.

Let us stress that, as usual, the most relevant process does not necessarily makes the strongest impact on observables. The impact also depends on  bare values of the inter-wire couplings omitted in Eq.~\eqref{inter}. Since   both $\hat{L}^{{\mathrm{sp}}}_{ij}  $ and $\hat{L}^{{\mathrm{sc}}}_{ij}  $ involve tunnelling between the $i^{{\mathrm{th}}} $ and $j^{{\mathrm{th}}} $ wires, their bare values can be much smaller than that of $\hat{L}^{{\mathrm{cdw}}}_{ij}  $ that involves only virtual processes. However, the bare values are totally irrelevant at $T=0$ so that the results based on the analysis of the RG dimensions  should survive at low enough temperatures, provided that the wires are not too widely separated and the most important processes involve only the next-neighbouring wires.  In the last section, we will come back to the situation when a direct inter-wire tunnelling is suppressed so that only CDW processes should be taken into account.

\section{Identical wires}
If all the wires are identical and packed into a 2D or 3D array, the labels $i$  are replaced by lattice vectors  ${\bm R}$ where ${\bm R}\subset {\cal L}$ and  ${\cal L}$ is the one- or two-dimensional lattice of wires. Assuming the interactions to be  translationally invariant ($V_{ij}\to V_{|{\bm R}-{\bm R}'}|$), Eq.~\eqref{Kmat} for the Luttinger matrix takes the following form:
\begin{align}\label{Klat}
&\sum_{{\bm R}_1,{\bm R}_2\subset{\cal L}}\,K_{{\bm R}-{\bm R}_1}\,V^{\theta}_{{\bm R}_1-{\bm R}_2}\,K_{{\bm R}_2-{\bm R}'}=V^{\varphi}_{{\bm R}-{\bm R}'}\,,
\end{align}
where ${\bm{r}}\equiv {\bm R}-{\bm R}'\subset{\cal L}$ and lengths are measured in units where the inter-wire distance is put to $1$.
  This
equation   is solved  via the Fourier transform:
\begin{equation}\label{FT}
\begin{aligned}
&K_{{\bm r}}=\int \frac{{\rm d}^dq}{(2\pi)^d}\,K_{\bm q}\,\e^{i{\bm q}{\bm r}}\,,\qquad
K_{\bm q}=\sqrt{\frac{V^{\varphi}_{\bm q}}{V^{\theta}_{\bm q}}}\,,\\ &V^{\theta/\varphi}_{\bm q}=V^{\theta/\varphi}_0+  \sum_{\bm r\ne\bm0 } \,U^{\theta/\varphi}_{\bm r}\,\e^{-i{\bm q}{\bm r}}\,.
\end{aligned}
\end{equation}
Here and below the momentum integration is carried out over a  Brillouin zone.
This results in the following expressions for the scaling dimensions \eqref{scal1} for the coupling between wires separated by the lattice vector ${\bm{r}}$:
\begin{eqnarray}\label{sc dim}
\begin{aligned}
\Delta_{{\bm r}}^{\rm cdw}&=2\int \frac{{\rm d}^dq}{(2\pi)^d}\,K_{\bm q}\left[1-\cos{\bm q}{\bm r}\right];\\
\Delta_{{\bm r}}^{\rm sc}&=2\,\int\frac{{\rm d}^dq}{(2\pi)^d}\,K_{\bm q}^{-1}\,\left[1-\cos{\bm q}{\bm r}\right]\,,\\
\Delta_{{\bm r}}^{\rm sp}&=\frac{1}{4}\left[\Delta_{{\bm r}}^{\rm cdw}+\Delta_{{\bm r}}^{\rm sc}\right]\,.
\end{aligned}
\end{eqnarray}
We have assumed above that the wires are arranged into a simple Bravais lattice. For a non-Bravais lattice,  the summation in matrix equation \eqref{Klat} should be carried out over all the sites in an elementary cell, with the appropriate changes to the Fourier-transformed solution.

\begin{figure*}\vskip 0.5 cm
\centering
\includegraphics[width=0.3 \linewidth]{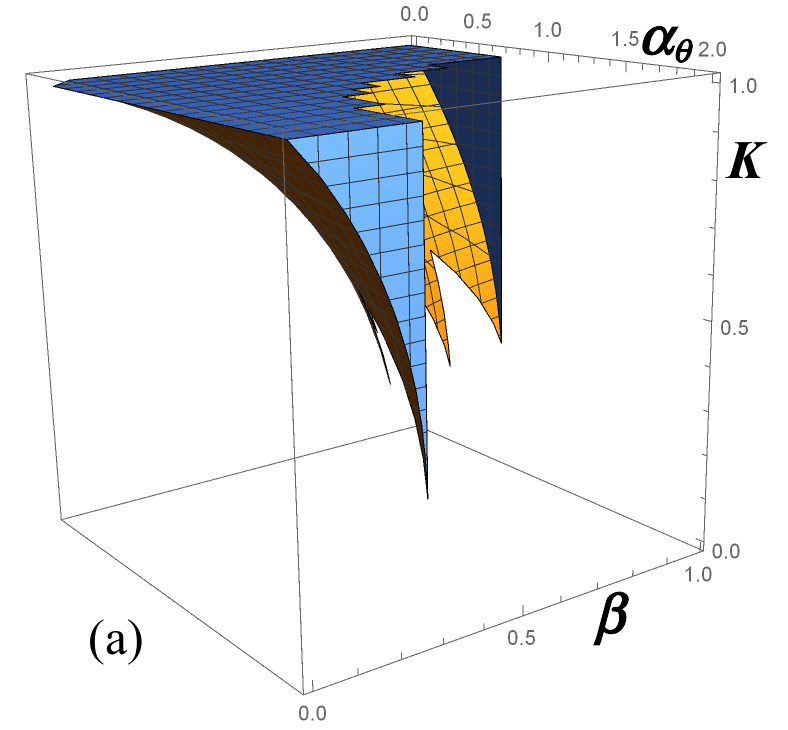}\qquad
\includegraphics[width=0.18 \linewidth]{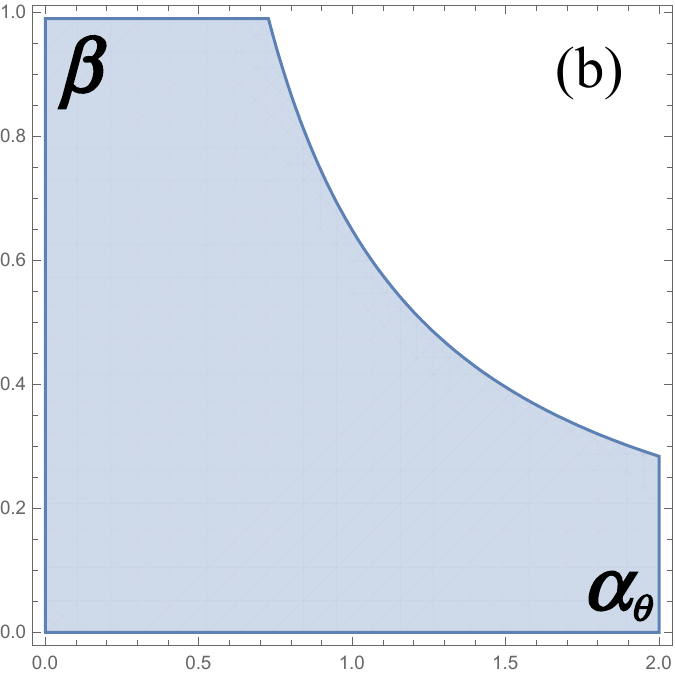}
\includegraphics[width=0.18 \linewidth]{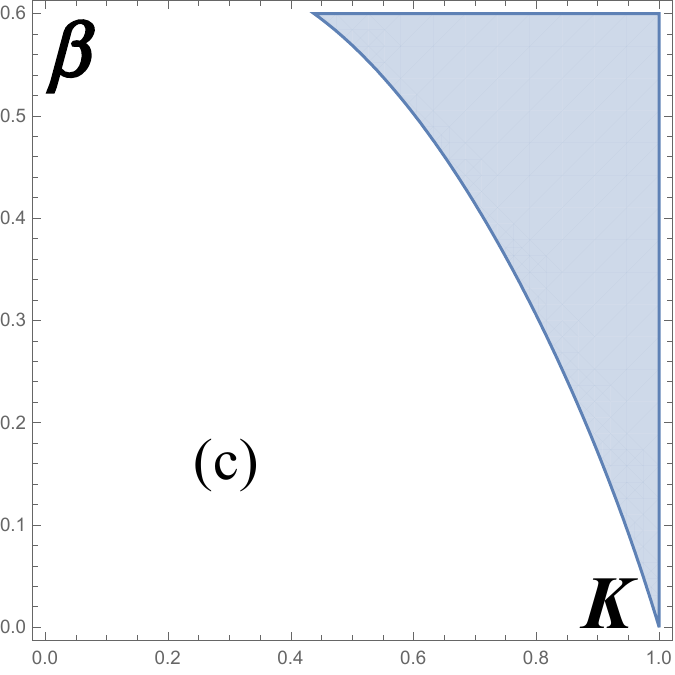}\qquad
\includegraphics[width=0.18 \linewidth]{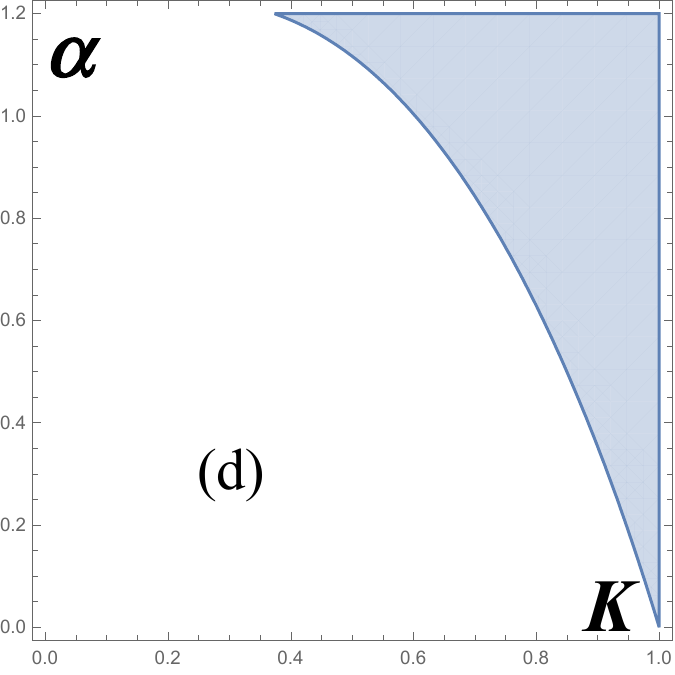}
\caption{Stability diagram of the SLL phase for a fermionic system with the intra-wire repulsion ({$K\leqslant1$}) without a direct inter-wire tunnelling. Here the strength of the density-density inter-wire interaction is parametrised by $\alpha_\theta$, and its range by  the screening parameter $\beta\equiv {\mathrm{e}}^{-\kappa} $  (with $\kappa^{-1} $ being the screening length measured in units with the inter-wire distance equal 1): (a) in $\alpha_{\theta}-\beta-K$ space; (b) in  $\alpha_{\theta}-\beta$ plane for a system without the intra-wire interaction ($K=1$); (c)  in $K-\beta$ plane for a system with $\alpha_{\theta}=1$; (d) in  $K-\alpha_{\theta}$ plane for a system with  $\beta  =0.5$.}
 \label{fig1}
 \end{figure*}

\section{Screened Coulomb  interaction}
In the absence of inter-wire interactions, when ${\mathsf{U}}=0$ in Eq.~\eqref{V}, the Luttinger matrix is diagonal, with all elements equal to $K$.  Assuming that  only  the nearest neighbouring wires are coupled for all the three perturbations, the scaling dimensions  \eqref{scal1} are reduced to
\begin{align}
\Delta_{0}^{\rm cdw}&=2K\,,& \Delta_{0}^{\rm sc}&=\frac{2}{K}\,,& \Delta_{0}^{\rm sp}&=\frac{1}{2}\left[K+\frac{1}{K}\right]\,.
\label{scal3}
\end{align}
Obviously, there is no value of $K$  for which all the scaling dimensions are above $2$  so that the SLL is unstable, at least in the absence of long-range interactions.
A weak short-range inter-wire interaction cannot stabilise the SLL phase since it gives only small corrections to the RG dimensions \eqref{scal3} which are never simultaneously close to $2$.

Let us consider the case of a weakly screened Coulomb interaction,  with $U^{\theta/\varphi }_{\bm{r}}= {\alpha_{\theta/\varphi}}\,e^{-\kappa r}/{r}$ in Eq.~\eqref{FT} and   $\kappa{\ll} 1$.  Then, we represent $V^{\theta/\varphi }_{\bm q}$ in Eq.~\eqref{FT} for $2D$ and $3D$ arrays as follows:
\begin{align}\label{u}
     \begin{aligned}V^{\theta/\varphi }_{\bm q}&=V_0^{\theta/\varphi }  \left[1 +\alpha_{\varphi/\theta}\,u_{\bm q}\right]\,,\\
        u_{\bm q}^{(2D)}&=-\ln\left[1-2\beta\,\cos q+\beta^2\right]\,,\quad \beta\equiv \e^{-\kappa}\,, \\u_{\bm q}^{(3D)}&=\sum_{{\bm n}\neq 0} \frac{\exp[-\kappa\sqrt{n_1^2+n_2^2}]}{\sqrt{n_1^2+n_2^2}}\,\e^{-i(q_1n_1+q_2n_2)}\,.
     \end{aligned}
\end{align}
The Fourier-transform of   Luttinger matrix ${\mathsf{K}}$, Eq.~\eqref{FT},  is expressed via $u_{\bm{q}}$ as
\begin{align}
K_{\bm q}&\equiv KQ({\bm q}),&Q({\bm q})&=\sqrt{ \frac{1+\alpha_{\varphi}\,u_{\bm q}} {1+\alpha_{\theta}\,u_{\bm q}}  }\,.
\label{Q}
\end{align}
The expression under the square root could become negative for some $q$ if any of the inter-wire interaction strength, $\alpha_\varphi $ or $\alpha_\theta$, exceeds $\alpha_{\rm WB}\equiv \left(2\ln(1+\beta)\right)^{-1}$, the boundary of the Wentzel-Bardeen \cite{WB} instability, typical for any multi-channel system. The standard LL approach is not valid there so that we assume that both interaction strength are bounded from above.

The scaling dimensions  for the next-neighbours coupling  are then expressed as
\begin{equation}\label{D3}
\Delta^{\rm cdw}=2K\langle Q_q\rangle\,,\quad \Delta^{\rm sc}=2K^{-1} \langle Q_q^{-1} \rangle\,,
\end{equation}
where the angular brackets are defined by
\begin{equation*}
\langle f\rangle\equiv \left\{\begin{array}{cc}\displaystyle
                         \int\limits_0^{2\pi}\frac{{\rm d}q}{2\pi}f_{q}\,(1-\cos q)\,, & 2D; \\[15pt]
                     \displaystyle    \int\limits_0^{2\pi}\frac{{\rm d}q_1}{2\pi}\frac{{\rm d}q_2}{2\pi}f_{\bm q}\,(1-\cos q_1)\,, & 3D\,.
                       \end{array}\right.
\end{equation*}
 The inter-wire interaction opens up a potential region of stability as, in  contrast to the case of isolated channels,  the inequalities $\Delta^{\rm cdw}>2$ and $\Delta^{\rm sc}>2$ can be both satisfied provided that $\langle Q_q\rangle^{-1} <K<\langle Q_q^{-1} \rangle $. The third condition of stability, $\Delta^{{\mathrm{sp}}}=\frac{1}{4}({\Delta^{{\mathrm{cdw}}} +\Delta^{\mathrm{sc}}})>2 $ is more stringent, as it can be satisfied simultaneously with the two previous ones only if
\begin{equation}
\langle Q_q\rangle\,\langle Q^{-1}_{q}\rangle > 3\,,
\label{cond1}
\end{equation}
and either $\langle{Q_q}\rangle^{-1}<K<K_+$ or $K_-<K<\langle{Q_q^{-1} }\rangle$, where $ K_\pm= 2\langle{Q_q}\rangle^{-1}\big[{ 1\pm \big({1{-}\frac{1}{4}\langle{Q_q}\rangle^{-1}\langle{Q_q^{-1}}\rangle}\big)^{\frac{1}{2}}  }\big]  $.
  For repulsive fermions, $K<1$, this can only happen when $\langle{Q_q}\rangle>1$.
When we study numerically these stability conditions, we find that for 2D and 3D packing there is no stability region of the fermion SLL model. Note in passing that such a region may exist for $K>1$. However, for  this case most easily realisable for multi-channel liquid of ultra-cold bosons the model with a long-range inter-wire interaction is not realistic.

\section{`Coulomb-blockade' interaction}
  Now we consider an ultimate long-range interaction, independent on the inter-wire distance. The interaction matrices (in proper units) have the form:
\begin{align}
{\mathsf{V}}&=({1-\alpha})\,{\mathbb{1}} +\alpha\,{\mathsf{E}}\ & \Leftrightarrow&& V_{ij} &=\delta_{ij}+\alpha\left[1-\delta_{ij}\right]\,.
\label{Cb}
\end{align}
where all elements of matrix ${\mathsf{E}}$ are equal to unity. The coefficients $\alpha\to \alpha_{\theta,\,\varphi }$ for the  two types of  interaction in Eq.~\eqref{V},
All such matrices commute with each other.   Thus, the solution of Eq. (\ref{Kmat}) for the Luttinger matrix   is
\begin{equation*}
{\mathsf{K}}=  \mathsf{V}_{\varphi}^{1/2} \,\mathsf{V}_{\theta}^{-1/2} \!.
\end{equation*}
Assuming a finite number of wires ({$=N$}) and noticing that ${\mathsf{E}}^2=N\,{\mathsf{E}}$, one expresses ${\mathsf{V}}^{-1} $ and ${\mathsf{V}}^{1/2} $ as
\begin{align*}
{\mathsf{V}}^{-1}&=\frac{1}{1-\alpha}\left[{\mathbb{1}}-\frac{\alpha}{N{\alpha}+1-\alpha}\,{\mathsf{E}}\right]\,,\\
 {{\mathsf{V}}}^{\frac{1}{2}} &=\sqrt{1-\alpha}\left[{\mathbb{1}} +\frac{1}{N}\left(\sqrt{1+\frac{N{\alpha}}{1-\alpha}}-1\right)\,{\mathsf{E}}\right]\,.
\end{align*}
Using this expressions one   finds, with $c_{\varphi/\theta}=\frac{\alpha_{\varphi/\theta}}{1-\alpha_{\varphi/\theta}}$:
\begin{equation}
{\mathsf{K}}=K\,\sqrt{\frac{1-\alpha_{\varphi}}{1-\alpha_{\theta}}}\,\left[{ {\mathbb 1}}+\frac{1}{N}\left(\sqrt{\frac{Nc_{\varphi}+1}{Nc_{\theta}+1}}-1\right)\,{\mathsf{E}}\right].
\end{equation}
Here again both interaction strength are bounded from above, $\alpha_{\theta,\varphi }<1 $, by the Wentzel-Bardeen \cite{WB} instability.
 The inverse  matrix ${\mathsf{K}}^{-1} $ is obtained by swapping $\theta\leftrightarrow\varphi$.
Substituting these expressions into Eq.~(\ref{scal1}), the off-diagonal elements cancel out, and only the diagonal ones contribute  to the scaling dimensions:
\begin{equation}
\Delta^{\rm cdw}=2K\,Q\,,\quad \Delta^{\rm sc}=\frac{2}{K\,Q}\,,\quad Q=\sqrt{\frac{1-\alpha_{\varphi}}{1-\alpha_{\theta}}}\,.
\end{equation}
Thus, in contrast to the case of the long-range (screened) Coulomb inter-wire interaction, the infinite-range `Coulomb-blockade-type' interaction simply renormalises the effective Luttinger parameter ($K\to Q\,K$) and reproduces the results without inter-wire interactions, Eq.~(\ref{scal3}).
Therefore, when all the three perturbations, CDW, SC and SP, exist, there is no stable SLL-phase (where all the perturbations are irrelevant)  for arrays of 1D channels and for all reasonable forms of intra- and inter-wire repulsions.

\section{Suppressed inter-channel scattering}
Here we show that a stable SLL can only be  realised in a system without a direct inter-channel tunnelling. This can happen in a spin-gapped system where the single-particle tunnelling is suppressed or for a sufficiently large inter-wire distance where both single-particle and pair {SC} tunneling bare values are small and, therefore, can be neglected at not too low temperatures. In the latter case case we have to consider, alongside with the (long-range) Coulomb interaction,  the {CDW} perturbation only. It is reasonable to assume that the current-current inter-wire interaction is much weaker than the density-density one so that we  put $\alpha_{\varphi}=0$ in Eqs. (\ref{u}) and \eqref{Q}. We the find $\Delta^{{\mathrm{cdw}}} $, Eq.~\eqref{D3}, numerically and present our results for the 2D array in Fig.~1. The SLL stability region  is shown in Fig.~1a  in $\alpha_{\theta}-\beta-K$ space. The strength of the inter-wire interaction is bounded, $|\alpha_{\theta}|\leq\alpha_{\rm WB}\equiv \left(2\ln(1+\beta)\right)^{-1}$, because above the critical value $\alpha_{\rm WB}$ the  Wentzel-Bardeen \cite{WB} instability  occurs and this regime is beyond the applicability of our theory. Within the bounds, one  immediately notices a competition between intra- and inter-wire interactions. Indeed, as we illustrate in Fig.~1b for a case with no intra-wire interaction  ($K=1$), any weak inter-wire interaction immediately stabilises SLL. The graphs in Fig.~1c,d also illustrate that turning on the inter-wire interactions  stabilises the SLL phase even for $K<1$. Thus, the SLL phase can be in principle observed in the multi-channel array provided that the inter-wire tunnelling necessary for the SP and SC perturbations is suppressed.

\section{Conclusions}
After analysing all allowed perturbations (single-particle tunneling and two-particle charge-density wave and Josephson couplings) in a SLL consisting of identical channels/wires with no magnetic field applied, we have shown that no physically reasonable inter-wire interaction can stabilize the model, i.e. support 'confinement' of particles and their pairs and the 'sliding' phase at the same moment. The situation is qualitatively similar in 2D- and 3D-packed arrays. The only possibility for such a description to be valid at the lowest temperature is the suppression of inter-wire hybridisation that can be achieved when either system is spin-gapped (attraction in the spin sector) or magnetic field is applied.

\begin{acknowledgments}
VF and IVY acknowledge the hospitality of the Center for Theoretical Physics of Complex Systems, Daejeon, South Korea, where this project has been started. IVY's research was funded by the Leverhulme Trust Research Project Grant RPG-2016-044.
\end{acknowledgments}

\end{document}